\documentclass[
12pt,tightenlines, amsmath, amssymb, nofootinbib, prd,
superscriptaddress, showpacs, preprintnumbers]{revtex4}
\newcommand{\beq}{\begin{equation}}
\newcommand{\eeq}{\end{equation}}
\newcommand{\ba}{\begin{array}{ccc}}
\newcommand{\ea}{\end{array}}
\newcommand{\nn}{\nonumber}
 \renewcommand{\d}{\partial}
\def\bea{\begin{eqnarray}}
\def\eea{\end{eqnarray}}

\def\<{\langle}
\def\>{\rangle}

\usepackage{graphicx}
\usepackage{amsmath}
\usepackage{amssymb}
\usepackage{psfrag}
\usepackage{epsfig}
\begin{document}

\title{\Large{{\bf Is Schwinger Model at Finite Density a Crystal?}}}
 \affiliation{Department of Physics and
Astronomy, University of British Columbia, Vancouver, BC, Canada,
V6T 1Z1}
 \author{ Max~A.~Metlitski}
 \email{mmetlits@phas.ubc.ca}
 \affiliation{Department of Physics and
Astronomy, University of British Columbia, Vancouver, BC, Canada,
V6T 1Z1}
\date{\today }

\vfill
\begin{abstract}
It has been believed since the paper by Fischler, Kogut and
Susskind\cite{Susskind} that in $QED_2$ at finite charge density the
chiral condensate exhibits a spatially inhomogeneous, oscillating
behaviour. In this paper we demonstrate that this inhomogeneity is
due to unphysical explicit breaking of the translational invariance
by a
uniform background charge density. 
Moreover, we investigate in the context of a simple statistical
model what happens if the neutralizing background is composed
instead of heavy, but dynamical, particles. We find that in contrast
to the standard picture of \cite{Susskind}, the chiral condensate
will not exhibit coherent oscillations on large distance scales,
unless the heavy neutralizing particles themselves form a crystal
and the density is high.
\end{abstract}

\vfill

\maketitle

\section{Introduction}

Over the past years there has been a lot of interest in the phase
diagram of $QCD$ at finite temperature and baryon density. The phase
diagram would provide one with the answer to the childish question,
``What happens to matter when you heat it up or squeeze it?"
This question is relevant for the analysis of such extreme natural
environments as the early universe or the dense interiors of neutron
stars. The interest in the phase diagram of $QCD$ has in turn
sparked studies of models of $QCD$ in non-trivial environments.

Two dimensional Quantum Electrodynamics, $QED_2$, commonly referred
to as the Schwinger model, has served as a playground for $QCD$
theorists for many years. In the massless limit this model is
exactly solvable and displays many features similar to $QCD$, most
notably the generation of mass gap and the appearance of chiral
condensate. The chiral condensate in this case is a manifestation of
explicit chiral symmetry breaking by the axial anomaly and is
generated in sectors with topological charge $\pm1$.

The first study of the Schwinger model at finite density has been
performed in \cite{Susskind}. It must be noted that in this case we
are talking about electric charge density, which unlike the baryon
number density, is the zeroth component of a current associated with
a gauged rather than global symmetry. Thus, to study  $1$-flavour
Schwinger model at finite density one needs to introduce background
charge in order to satisfy the Gauss law. The natural choice for
such a background that was adopted in \cite{Susskind} is just a
finite external charge uniformly smeared along the spatial
direction.

The massless Schwinger model remains exactly solvable at finite
density. One of its most surprising features is that once an
arbitrary small charge density is introduced, the chiral condensate
is no longer spatially uniform, but instead supports a plane wave
structure\cite{Susskind,Kao},
\beq \label{psiwave}\langle \bar{\psi} \psi(x)\rangle = \langle
\bar{\psi} \psi\rangle_0 \cos(2 \pi \rho x + \theta) \eeq where
$\rho$ is the number density, $\theta$ is the topological angle of
$QED_2$ and $\langle \bar{\psi} \psi\rangle_0$ denotes the chiral
condensate at zero temperature, density and $\theta$ parameter.
Thus, the chiral condensate experiences oscillations with the period
given by inverse number density. On the other hand, as long as the
quark mass is vanishing the fermion density itself is uniform. So
the model does not develop a conventional crystal but rather a
``chiral crystal".

Once a finite quark mass $m$ is introduced, the non-uniformity of
the chiral condensate is translated into a non-uniformity of fermion
density and to leading order in $m/e$\cite{Susskind}, \beq
\label{nwave}\langle \bar{\psi} \gamma^0 \psi(x) \rangle \approx
\rho \left(1+\frac{4 \pi m}{\omega^2} \langle\bar{\psi}\psi\rangle_0
\cos(2 \pi \rho x + \theta)\right)\eeq where $\omega = e/\sqrt{\pi}$
and $e$ is the gauge coupling constant.

In this paper we will argue that the above conventional picture of
oscillating chiral condensate (\ref{psiwave}) is unphysical, being a
consequence of the introduction of an external static, neutralizing
charge density.

There have been a number of studies of the Schwinger model at finite
density both in the Hamiltonian\cite{Kao,Nagy} and path-integral
formalism\cite{Schaposnik} since the work \cite{Susskind}, all of
which have confirmed the oscillations of the chiral condensate.
However, the ultimate reason for the breaking of the translational
invariance in this model remains somewhat unclear. Indeed, one
generally cannot spontaneously break continuous symmetries in $1+1$
dimensions. Moreover, the ground state of the Schwinger model at
finite density is unique (once the topological $\theta$ angle is
fixed) and the lowest lying excitations are separated by a mass gap
$\omega$.

Yet, it is important to understand the precise reason for this
phenomenon not only because it is curious by itself, but also as
similar behaviour of the chiral condensate occurs in a large number
of other models. In particular, both the $2d$ Gross-Neveau model and
the two dimensional $QCD$ in the large $N_c$ limit are believed to
exhibit exactly the same spatial oscillations of the chiral
condensate at finite density\cite{Thies}. The period of oscillations
is again given precisely by the inverse fermion density. The
stability of the chiral crystal against quantum fluctuations in
these models is argued on the basis of the large $N_c$ limit: once
$N_c = \infty$ one is allowed to circumvent the Mermin-Wagner
theorem. Moreover, four-dimensional dense $QCD$ in the large $N_c$
limit is also expected to support a periodically modulated chiral
condensate\cite{Rubakov}. Thus, it would be valuable to first fully
understand the origin of translational symmetry breaking in $QED_2$,
which is considerably simpler than the above zoo of models.

We demonstrate, that the reason for this phenomenon is the presence
of the background charge density, which leads to the inability to
simultaneously maintain invariance under translational and large
gauge transformations. Alternatively, in the path integral language,
translational invariance is violated by sectors with a finite
topological charge.  These findings naturally explain the particular
form of the chiral condensate at finite density and provide a more
conclusive explanation for the loss of translational invariance than
those present in the literature. We show that although both the
chiral and translational symmetries are explicitly broken at finite
density, in the massless limit a linear combination of them remains
intact, which implies, \beq \langle O(x) \rangle = e^{-\pi i q \rho
x} \langle O(0) \rangle\eeq where $O(x)$ is an arbitrary local
operator and $q$ is the axial charge of $O$.

Having thoroughly understood the reason why the uniform background
density leads to explicit breaking of translational symmetry we ask
the following question. Should we consider such a theory completely
sick? More precisely, does the theory with the uniform background
charge ever correctly model a theory where the neutralizing charge
is heavy, but dynamical. Clearly, any theory where the
neutralization of charge is performed solely by dynamical fields
will not exhibit explicit breaking of translational invariance.
Moreover, in the absence of ``special arrangements" such as $N_c =
\infty$ the translational invariance will not be broken in two
dimensions spontaneously either. However, relics of the chiral
crystal might remain intact on some finite, but large, distance
scale.

To answer the above question we consider a system where the
neutralizing charge is modeled by dynamical classical particles of
integer charge. We expect that this model corresponds to a theory
where one fermion species is massless and the other is very heavy
(of mass $M$), in the regime $T \ll M$, $e \ll M$, where $T$ is the
temperature, provided that $T$ is sufficiently large that the
quantum effects for the heavy particles can be neglected. 
We integrate out the light degrees of freedom
(photons and massless fermions) and are left with a classical
statistical mechanics model for the heavy degrees of freedom. These
heavy degrees of freedom have a size of roughly $1/\omega$, interact
via a Yukawa potential and should probably be identified with
$B$-like mesons, consisting of one light and one heavy quark.

We find that the chiral condensate in this model will not reproduce
the standard picture of \cite{Susskind} (see eq. (\ref{psiwave})),
which exhibits for arbitrary density spatial oscillations with a
density independent amplitude. Instead, the form of the condensate
will depend crucially on the density and on the emergent dynamics of
the mesons. In the regime where the model is tractable (i.e. when
the mesons form a weakly interacting gas), the chiral condensate
does not reproduce any of the features of eq. (\ref{psiwave}).
Instead, in the dilute limit $\rho \ll \omega$, the chiral
condensate is uniform and its magnitude decreases slightly with
density. The correlator
$\langle\bar{\psi}\frac{1+\gamma^5}{2}\psi(x)
\bar{\psi}\frac{1-\gamma^5}{2} \psi(0)\rangle$ does not experience
any oscillations. In the high density limit, $\rho \gg \omega$ the
chiral condensate decreases exponentially with density. The
correlator $\langle\bar{\psi}\frac{1+\gamma^5}{2}\psi(x)
\bar{\psi}\frac{1-\gamma^5}{2} \psi(0)\rangle$ exhibits oscillations
with period $\rho^{-1}$ on distance scales $x \ll \omega^{-1}$,
which, however, disappear for $x \gg \omega^{-1}$. These
oscillations on short distances are the only visible remnants of the
chiral crystal in the gaseous regime.

Thus, we shall argue that the chiral condensate has a chance to
reproduce the plane wave behaviour (\ref{psiwave}) on sufficiently
large distance scales only if the density $\rho \gg \omega$ and the
heavy degrees of freedom themselves are close to crystallization.
Unless these specific conditions are met, the uniform background
charge approximation is inapplicable and the form of the chiral
condensate (\ref{psiwave}) found in \cite{Susskind,Kao,Schaposnik}
is unphysical.


\section{What's Non-Uniform in a Uniform Background Density}

This section is devoted to a detailed analysis of the reason for the
appearance of the chiral crystal in a model with a uniform
background density.  The literature on this subject generally
supports the following argument present in the original
paper\cite{Susskind}. If the spatial manifold is an infinite line
${\mathbb R}$, one prefers not to introduce a background charge
distribution that stretches across whole of ${\mathbb R}$ to avoid
infra-red difficulties. Instead, one chooses a background charge
density to be uniform in a certain finite region of the real line
(say $-L  < x < L$) and zero everywhere else. Once all the
calculations are done one takes $L \to \infty$. Then the ``small"
explicit breaking of translational symmetry present in the form of
the endpoints of the charge distribution is carried by the
long-range Coulomb forces across the whole system and leads to the
chiral crystal structure (\ref{psiwave}).

In principal, the above invocation of the long-range forces allows
one to circumvent the general theorems on the lack of spontaneous
symmetry breaking in $1+1$ dimensions. However, the above argument
can no longer be directly applied once the spatial manifold is
compactified to a circle with the background charge uniformly
smeared along its length, apparently removing the ``endpoints" of
the charge distribution. We adopt precisely such a compactification
of the spatial coordinate in what follows.

Moreover, let us compare the situation to Schwinger model at zero
density, where one observes breaking of the chiral symmetry. The
modern philosophy regarding the origin of this phenomenon is that
chiral symmetry is locally explicitly broken by the axial anomaly.
Globally, one cannot simultaneously maintain invariance of the
theory under chiral and large gauge transformations. Translating the
last statement into the path integral formalism: axial charge is not
conserved in non-trivial topological sectors.

We now show that a similar picture holds for the breaking of
translational invariance in Schwinger model at finite density.

\subsection{Hamiltonian Formalism}
We start with the Lagrangian for $QED_{1+1}$, \beq L =
-\frac{1}{4}F_{\mu \nu} F^{\mu \nu} + \bar{\psi} i
\gamma^{\mu}D_{\mu}\psi - m \bar{\psi} \psi, \quad D_{\mu} =
\d_{\mu} - i e A_{\mu}\eeq Local $U(1)$ symmetry of the theory takes
the form, \beq \label{U1} U(1):\, \psi(x) \to e^{i \alpha(x)}
\psi(x), \quad A_{\mu}(x) \to A_{\mu}(x) + \frac{1}{e}\d_{\mu}
\alpha(x)\eeq For the moment we work in Minkowski space, with the
conventions $\gamma^5 = \gamma^0 \gamma^1$, $\epsilon^{01} = 1$. For
definiteness, we take the spatial manifold to be a circle of length
$L_1$ and pick the gauge where all fields obey periodic boundary
conditions on this circle.

The energy momentum tensor for this theory is, \beq T^{\mu \nu} =
\bar{\psi} i \gamma^{\mu}D^{\nu} \psi - F^{\mu
\lambda}F^{\nu}_{\,\,\lambda} - g^{\mu \nu} L \eeq We have not
symmetrized $T^{\mu \nu}$ as it is not essential for our purposes.

Now let us couple the theory to a conserved external current
$j_{ext}^{\mu}(x)$, such that $\d_{\mu} j^{\mu}_{ext} = 0$.  The
Lagrangian becomes, \beq L_j = L + j_{ext}^{\mu} A_{\mu} \eeq Once
this term is added, the energy momentum tensor satisfies, \beq
\label{T1} \d_{\mu} T^{\mu \nu} = F^{\nu \lambda} j_{\lambda\,\,
ext}\eeq Clearly, an external current violates the conservation of
energy and momentum. Now, let us take $j^{\mu}$ to represent a
uniform, neutralizing charge density, \beq j^{\mu}_{ext} = (-e\,
\rho, 0)\eeq  so that $\rho = N/L_1$, where $N$ is the total
dynamical charge.

The equation (\ref{T1}) becomes, \bea \d_{\mu}T^{\mu 0} &=& 0\\
\d_{\mu} T^{\mu 1} &=& - e\, \rho \,F \eea where $F = F_{01}$, is
the electric field. Thus, the uniform background charge density
explicitly breaks
 spatial but not temporal translational invariance. In
particular, defining the total momentum operator, \beq P = \int dx^1
\, T^{01} \eeq we obtain, \beq \label{ddtP} \frac{d}{dt} P = - e\,
\rho\, \int dx^1 \, F \eeq If we integrate equation (\ref{ddtP})
over time, \beq \label{DP} \Delta P = - e \, \rho \int d^2 x \,
F\eeq We recognize the integral on the righthand side of eq.
(\ref{DP}) as the topological charge of the theory.

 Thus, translational invariance
is broken both locally and globally. One could try to redefine the
energy momentum tensor as, \beq \hat{T}^{\mu 1} = T^{\mu 1} + e
\,\rho\, \epsilon^{\mu \nu} A_{\nu} \eeq and likewise the total
momentum \beq \hat{P} = P + e\,\rho \int dx^1 A_1 \eeq so that \beq
\d_{\mu} \hat{T}^{\mu 1} = 0, \quad \quad \frac{d}{dt} \hat{P} = 0
\eeq However, the local current $\hat{T}^{\mu 1}$ is clearly not a
gauge invariant operator. The global object $\hat{P}$ is invariant
under ``small" gauge transformations characterized by $\alpha(L_1) =
\alpha(0)$, where $\alpha(x)$ is the transformation parameter of eq.
(\ref{U1}). However, $\hat{P}$ is not invariant under large gauge
transformations $U$, \beq \label{large} U \psi(x^1) U^{\dagger} =
e^{2 \pi i x^1/L_1}  \psi(x^1),\quad  U A_1 U^{\dagger} = A_1 +
\frac{2 \pi}{e L_1}\eeq whereby\beq U \hat{P}U^{\dagger} = \hat{P} +
2 \pi  \rho\eeq

Thus, at finite background charge density, we cannot simultaneously
preserve the invariance of the theory under both translational and
large gauge transformations. The usual procedure, at least at zero
density, is to formulate the theory in a way, which preserves the
latter symmetry and to constraint oneself to states in the Hilbert
space satisfying, \beq U |\theta\rangle = e^{i \theta}
|\theta\rangle\eeq Then a finite translation with the operator
$\hat{P}$ will take us out of the gauge invariant Hilbert space and
into a different $\theta$-vacuum: \beq e^{i \hat{P} a}
|\theta\rangle = |\theta + 2 \pi \rho a\rangle\eeq Thus, for any
local operator $O(x)$, \beq \label{thetax}\langle O(x)
\rangle_{\theta} = \langle O(0) \rangle_{\theta+2 \pi \rho x}\eeq
This interplay between the $\theta$ angle and the loss of
translational invariance is clear from the expressions for the
chiral condensate and baryon density (\ref{psiwave}),(\ref{nwave}).
We would like to point out that we have not anywhere used the fact
that our dynamical matter is fermionic. Thus, eq. (\ref{thetax})
would remain valid in a theory with any dynamical matter fields
neutralized by a uniform background charge density.

Note that a lattice subgroup of the translational group remains
unbroken. Indeed, the operator, \beq S = e^{i \, \hat{P} \,
\rho^{-1}}\eeq is invariant under the transformation (\ref{large}).
But $S$ is an operator that performs translations by a distance $a =
\rho^{-1}$ - the average charge spacing. Thus, our theory will
respect this symmetry as can be explicitly seen from
(\ref{psiwave}),(\ref{nwave}).

The above discussion is precisely analogous to the philosophy behind
the breaking of axial symmetry in $QED_{1+1}$. Recall that the gauge
invariant axial current, $j^{\mu5} = \bar{\psi} \gamma^{\mu}
\gamma^5 \psi$ suffers from an anomaly, \beq \label{anom} \d_{\mu}
j^{\mu5} = -\frac{e}{2 \pi} \epsilon^{\mu \nu} F_{\mu \nu} + 2 i m
\bar{\psi} \gamma^5 \psi(x)\eeq Equation (\ref{anom}) is an operator
identity and should not be affected by infra-red effects such as
temperature or finite density.  Let us define the following current,
\beq l^{\mu} = T^{\mu 1} - \pi \rho j^{\mu5} \eeq Observe, that
$l^{\mu}$ is a gauge invariant operator, satisfying, \beq
\label{dlm}\d_{\mu} l^{\mu} = -2 \pi i \, m \, \rho\, \bar{\psi}
\gamma^5 \psi\eeq So in the massless limit $m = 0$, \beq \d_{\mu}
l^{\mu} = 0\eeq Thus, at finite density, both the axial and the
translational symmetries are broken. However, in the massless limit,
a linear combination of them remains intact. Defining the global
charge, \beq \label{Q} Q = \int dx^1 \, l^0(x) = P - \pi \rho
Q^5,\quad \quad \frac{d}{dt} Q = 0\eeq The conservation of $Q$
dictates the structure of all ``non-uniformities" provided that the
symmetry associated with the conservation of $Q$ is not
spontaneously broken. Consider an arbitrary local operator of axial
charge $q$, \beq [Q^5, O(x)] = q O(x)\eeq Then, \beq \langle O(x)
\rangle = \langle \Omega |e^{-i Q a} O(x) e^{i Q a}| \Omega\rangle =
\langle \Omega |e^{\pi i \rho a Q^5} O(x+a) e^{-\pi i \rho a Q^5}|
\Omega\rangle = e^{\pi i q \rho a } \langle O(x+a) \rangle\eeq

\beq \label{theor}\langle O(x+a) \rangle = e^{-\pi i q \rho a}
\langle O(x) \rangle \eeq \vspace{0.25cm}

In particular, for the fermion bilinear $\bar{\psi}
\frac{1\pm\gamma^5}{2} \psi$, $q = \mp 2$ and, \beq \langle
\bar{\psi} \frac{1\pm\gamma^5}{2} \psi(x)\rangle =e^{\pm 2 \pi i
\rho x} \langle \bar{\psi} \frac{1\pm\gamma^5}{2} \psi(0)
\rangle\eeq Thus, we see that the plane wave behaviour of the chiral
condensate follows immediately from the structure of the theory. On
the other hand, the density operator $\bar{\psi} \gamma^0 \psi$ has
$q = 0$ and, therefore, does not display any non-uniformity in the
massless limit. Thus, the equation (\ref{theor}) is in agreement
with the explicit calculations of \cite{Susskind,Kao,Schaposnik}.

Once the quark mass $m$ is non-vanishing the conservation of the
current $l^{\mu}$ is explicitly broken (\ref{dlm}). Therefore,
averages of local operators no longer need to satisfy the formula
(\ref{theor}). For instance, the fermion density $\langle \bar{\psi}
\gamma^0 \psi(x) \rangle$ becomes non-uniform as can be seen from
eq. (\ref{nwave}).

Before we conclude this section, we note that beside Schwinger
model, both two dimensional chiral Gross-Neveu model and $QCD_2$ are
believed to display the structure (\ref{theor}) in the large $N$
limit\cite{Thies}. In these theories both axial and translational
symmetries are spontaneously broken, but the linear combination
(\ref{Q}) remains preserved by the ground state. Thus, the resulting
picture is the same as in Schwinger model, but the formal reason for
the appearance of a chiral crystal is very different. In Schwinger
model, as we have shown, translational and axial symmetries are
broken explicitly (by background charge density and by chiral
anomaly). In the Gross-Neveau model and $QCD_2$ these symmetries are
broken spontaneously, with the theorems on the absence of
spontaneous symmetry breaking in two dimensions circumvented due to
$N = \infty$.

\subsection{Path-Integral Formalism}

It is instructive to understand in parallel how translational
symmetry breaking is realized in the path integral formalism.

We go to Euclidean space with, \beq L_E = \frac{1}{4}
F_{\mu\nu}F_{\mu \nu} + \bar{\psi}\gamma_{\mu}D_{\mu}\psi + m
\bar{\psi} \psi\eeq In our notations $\gamma_1 \gamma_2 = i
\gamma_5$ , $\epsilon_{12} = 1$.

We take the space-time to be a torus with $0 \leq x_1 \leq L_1$, $0
\leq x_2 \leq L_2$. Physically, $L_2 = \beta = T^{-1}$ is the
inverse temperature. Gauge fields on a torus fall apart into sectors
classified by the topological charge, \beq n = \frac{e}{2\pi} \int
d^2 x \, F \eeq where $F = F_{12}$. In a general topological sector,
the gauge and fermion fields are not strictly periodic, but satisfy,
\bea \psi(x_1,L_2) = V_1(x_1) \psi(x_1,0),\quad
A_{\mu}(x_1,L_2) = A_{\mu}(x_1,0) - \frac{i}{e} \d_{\mu} V_1(x_1) V_1^{\dagger}(x_1)\label{bc1}\\
\psi(L_1,x_2) = V_2(x_2) \psi(0,x_2),\quad A_{\mu}(L_1,x_2) =
A_{\mu}(0,x_2) - \frac{i}{e} \d_{\mu} V_2(x_2)
V_2^{\dagger}(x_2)\label{bc2}\eea with $V_1$, $V_2$ satisfying the
consistency conditions, \beq V_1(0) V_2(L_2) = V_2(0) V_1(L_1)\eeq
The transition functions $V_1, V_2$ in turn determine the
topological charge $n$.

For each $n$, we have some gauge freedom in choosing $V_1, V_2$. For
instance, one choice is to have fermions anti-periodic in the
temporal ($x_2$) direction, so that, \beq \label{anti} V_1(x_1) =
-1, \quad
V_2(x_2)  = e^{2 \pi i n x_2/L_2}\eeq 

Let us recall that an external heavy static particle is inserted
into the theory in the form of a temporal Wilson loop. For instance,
the partition function in the background of $m$ static charges
located at points $\{x_i\}$ and with charges $\{e \, p_i\}$, $p_i
\in {\mathbb Z}$ is, \beq Z = \int \, {\cal D} A {\cal D} \bar{\psi}
{\cal D} \psi \, \prod_{i = 1}^{m} W(x_i,p_i)\, e^{-S} e^{i n \theta
}\eeq with \beq \label{loop} W(x,p) = \left(-V_1(x_1)^{-1} e^{i e
\int A_{2}(x,\tau) d\tau}\right)^{p}\eeq We have inserted the
prefactor $(-1)^p$, so that in a gauge where $V_1(x_1)=-1$, $W(x,p)$
reduces to the standard form, \beq W(x,p) = \exp(i p\,e\int
A_2(x,\tau) d\tau)\eeq Expression (\ref{loop}) is completely gauge
invariant and geometrically is the transport with respect to $A$
along a temporal cycle, taken in representation $p$ of the $U(1)$
group.

It is clear that once $p$ ceases to be an integer the expression
(\ref{loop}) for $W(x,p)$ becomes ambiguous (the only
representations of the $U(1)$ group are integral). This is not
surprising - it is precisely for this reason that the existence of
monopoles enforces quantization of electric charge\footnote{However,
the question of confinement of fractional charges in massless and
massive Schwinger model has been discussed for
ages\cite{SusskindCol,Gordon,Klebanov}. This question has to be
understood in the sense $e^{i e p \oint_{\gamma} A_{\mu} dx_{\mu}}
\equiv e^{i e p \int_D d^2x F}$ where $D$ is the region of our
manifold, such that $\d D = \gamma$. The Wilson loop with the
fractional charge is well defined only once we also choose $D$ and
is not independent of this choice.}. In the present case the role of
monopoles is played by $2d$ instantons. Similarly, it is problematic
to generalize the prefactor in $W$ involving the transition
functions $V$ to a continuous background charge distribution in a
manifestly gauge invariant manner.

We may still attempt to take the limit of a continuous charge
distribution in a particular gauge. The choice $V_1(x_1) = -1$ seems
to be most suited for this purpose.  As noted above, as long as we
are working with integral charges in this gauge, the transition
functions drop out of expression (\ref{loop}). Now we can take the
``continuum" limit, \beq \label{cont} Z = \int \, {\cal D} A {\cal
D}\bar{\psi} {\cal D} \psi \, e^{i e \int d^2 x\,j^{ext}_{2} A_2}\,
e^{-S} e^{i n \theta}\eeq where $j^{ext}_2(x_1)$ is the static
background charge density and (anti)periodic temporal boundary
conditions on (fermions) gauge fields are assumed from here on.
Expression (\ref{cont}) is not invariant under gauge
transformations, which change these boundary conditions.

Now let us address the question of translational symmetry breaking.
First, to understand the root of the problem consider a fractional
charge $p$ situated at $x_1 = L_1^-$ and move it across the
artificial cut at $x_1 = 0 \sim L_1$ to $x_1 = 0^+$. Observe, \beq
W(L_1,p) = e^{i e p \int A_2(L_1, \tau) d\tau} = e^{i e p \int
A_2(0, \tau) d\tau}e^{p \int \d_2 V_2(\tau) V^{\dagger}_2(\tau)} =
e^{2 \pi i p n} W(0,p)\eeq Thus, the cut on the torus is visible to
a fractional charge and invisible to an integer charge. Of course,
there is nothing new in this result. However, it is precisely this
fact that leads to translational symmetry breaking.

Indeed, take an arbitrary local operator $O(x)$ and pick $a > 0$
such that $0 < x, x+a < L_1$. Let us compute the expectation value
of $O(x)$ in the background of the charge distribution
$j^{ext}_{2}$. Then, \beq \langle O(x+a) \rangle = \frac{1}{Z} \int
\, {\cal D} A {\cal D}\bar{\psi} {\cal D} \psi \, O(x+a) e^{i e \int
d^2 x\,j^{ext}_{2} A_2}\, e^{-S} e^{i n \theta}\eeq We make the
following change of variables, \bea \psi'(x_1,x_2) &=&
\psi(x_1+a,x_2), \, A'_{\mu}(x_1,x_2) =
A_{\mu}(x_1+a,x_2),\,0 < x_1 < L_1 - a\\
\psi'(x_1,x_2) &=& V_2(x_2)\psi(x_1+a-L_1,x_2),\, L_1-a < x_1 < L_1
\\\,A'_{\mu}(x_1,x_2) &=& A_{\mu}(x_1+a-L_1,x_2) -
\frac{i}{e}\d_{\mu}V_2(x_2)V^{\dagger}_2(x_2),\, L_1-a < x_1 <
L_1\eea It is easy to see that $\psi'$, $A'$ obey the same boundary
conditions as original variables $\psi$, $A$. Thus, \beq \langle
O(x+a) \rangle = \frac{1}{Z} \int \, {\cal D} A' {\cal D}\bar{\psi'}
{\cal D} \psi' \, O'(x) e^{-2\pi i n \int_0^a dx_1 j^{ext}_2(x_1)}
e^{i e \int d^2 x\,j'^{ext}_{2} A'_2}\, e^{-S} e^{i n \theta}\eeq
where \beq j'^{ext}_2(x_1) = \left\{\begin{array}{ll}
j^{ext}_2(x_1+a) & 0 < x_1 < L_1 - a\\j^{ext}_2(x_1+a-L_1) &  L_1 -a
< x_1 < L_1\end{array}\right.\eeq is just the properly shifted
background charge density. Thus, we get an extra factor, \beq
\label{factor} e^{-2\pi i n \int_0^a dx_1 j^{ext}_2(x_1)}\eeq
related to the amount of charge passing through the cut during our
translation. As long as this charge is an integer, the factor
(\ref{factor}) is unity and the cut is invisible.

Now suppose our background charge is uniformly smeared across the
spatial circle, $j^{ext}_2(x_1) = -\rho$, where $\rho =
\frac{N}{L_1}$ is the dynamical charge. The background charge
density itself is invariant under shifts along the circle,
$j'^{ext}_2 = j^{ext}_2$ and the factor (\ref{factor}) becomes,
$e^{2 \pi i n \rho a}$. Hence, \beq \langle O(x+a) \rangle_{\theta}
= \langle O(x)\rangle_{\theta + 2 \pi \rho a} \eeq in agreement with
(\ref{thetax}). Moreover, if we disentangle the contributions to
$\langle O(x) \rangle$ coming from distinct topological sectors,
\beq \label{On} \langle O(x) \rangle_n = e^{2 \pi i n \rho x}
\langle O(0) \rangle_n\eeq Thus, different topological sectors
contribute to $\langle O(x) \rangle$ with different ``harmonics."
One effect of (\ref{On}) is that only the topologically trivial $n =
0$ sector contributes to the partition function and the topological
susceptibility vanishes even if the quark mass $m$ is non-zero.

Finally, if the quark mass is vanishing then due to fermion zero
modes, operators with axial charge $q$ get a contribution only from
sectors with $n = -\frac{q}{2}$. Thus, for $m = 0$, \beq \langle
O(x) \rangle = e^{- \pi i q \rho x} \langle O(0) \rangle\eeq in
agreement with (\ref{theor}).

Before we conclude this section, we would like to note that in the
massless limit explicit calculations can be done for the theory
defined by (\ref{cont}) on a finite torus. These are summarized in
the appendix. In particular, \beq \label{osc}\langle
\bar{\psi}\frac{1+\gamma^5}{2}\psi(x)\rangle = \frac{1}{2} e^{i
\theta} e^{2 \pi i \rho x} \langle\bar{\psi}
\psi\rangle_{L_1,L_2}\eeq where $\langle\bar{\psi}
\psi\rangle_{L_1,L_2}$ is the chiral condensate at zero density and
$\theta$ parameter on the torus of dimensions $L_1, L_2$. The result
(\ref{osc}) generalizes the path-integral computation of
\cite{Schaposnik}, which was performed on an infinite Euclidean
space to the case of a finite torus, where all the infra-red
singularities are under complete control. Technically, the
oscillating factor $e^{2 \pi i \rho x}$ comes from averaging the
fermion zero mode over torons (constant parts of field $A_{\mu}(x) =
t_{\mu}$) in the presence of background charge.

\section{Dynamical Background}
\subsection{General Remarks}
In the previous section we saw that a uniform background charge
density explicitly breaks translational invariance. This effect may
seem to be rather unphysical, in particular as it has to do with the
Dirac string being visible to our background charge. But after all,
the uniform background density is typically taken to model some
heavy, but dynamical, particles. Once all fields are dynamical, it
is clear that translational symmetry is not explicitly broken.
However, one would like to ask whether any features of the chiral
crystal discussed in the previous section remain.

To answer the above question, we would like to analyze $QED_2$ with
two flavours. We take one fermion flavour to have charge $e$ and
vanishing mass and the other flavour to have charge $q e$, $q \in
{\mathbb N}$ and mass $M \gg e$. We want to analyze the problem at a
finite ``isospin" density, with the heavy fermions neutralizing the
light ones. We work at finite temperature $T$. We will treat the
problem in a ``Born-Oppenheimer" like approximation. Namely, we
first freeze the positions of heavy particles, treating them as
static external charges, and integrate over the light fermions and
gauge fields. For instance, the partition function of the system in
the background of $N$ external charges situated at points
$\{{x_i}\}$ is, \beq Z(x_1,..x_N) = \langle \prod_i W(x_i,
-q)\rangle_l \eeq where the subscript $l$ denotes integration over
the light degrees of freedom. We then promote the external charges
to dynamical degrees of freedom, treating them as classical
particles. For example, the full partition function takes the form,
\beq \label{eff}Z = \sum^{\infty}_{n=0} \frac{z^n}{n!} \int
dx_1..dx_n \langle \prod_i W(x_i, -q)\rangle_l \eeq Here $z =
\frac{1}{\lambda} e^{\beta (\mu-M)}$ is the activity, $\mu$ is the
chemical potential and $\lambda = (\frac{2 \pi}{M T})^\frac12$ is
the thermal wavelength. Similarly, the expectation value of some
operator $O$ involving light quark fields is, \beq \label{O}\langle
O \rangle = \frac{1}{Z} \sum^{\infty}_{n=0} \frac{z^n}{n!} \int
dx_1..dx_n \langle O\, \prod_i W(x_i, -q)\rangle_l \eeq We will
often use the notation, \beq \langle O\rangle_{\{x_i, q_i\}} =
Z^{-1}(x_1,..x_n)\langle O \prod_i W(x_i, q_i)\rangle_l \eeq

We shall shortly see that after integration over the light degrees
of freedom, the heavy fermions get dressed into meson like
particles, consisting (in terms of quantum numbers) of $q$ light and
one heavy quark. So the effective theory (\ref{eff}) should be
understood as describing classical dynamics of such mesons. We
expect such an approximation to be valid as long as $M \gg e, M \gg
T$ so that the heavy quark-antiquark pairs do not get excited either
virtually or thermally. Moreover, we need $T$ to be high enough that
the meson gas/liquid is in a classical rather than quantum regime.
In the dilute gas limit, we expect that the system can be treated
classically for $T \gg \frac{\rho^2}{2M}$.

As a first step to analyze the resulting system, we need to perform
the integration over the light fermions and gauge fields. We shall
work on a Euclidean torus of spatial and temporal lengths $L_1$,
$L_2 = T^{-1}$. The expectation value of a product of straight
temporal Wilson loops in the massless Schwinger model has been
computed in a number of works\cite{Gordon,Zahed}. The result is (see
the appendix for a sketch of the calculation), \bea \langle \prod_i
W(x_i, q_i)\rangle_l &=& \exp\left(-L_2\, \frac{1}{2}\sum_{i,j}q_i
q_j e^2 V(x_i - x_j)\right)\\\label{V}V(x) &=&
\frac{1}{L_1}\sum_{p_1} \frac{1}{p^2_1 + \omega^2}\, e^{i p_1 x}\,
\stackrel{L_1\to \infty}{=}\frac{1}{2\omega}e^{-\omega |x|}\eea
where $\omega = e/\sqrt{\pi}$ and $p_1 = 2 \pi m/L_1$, $m \in Z$.
Thus, our heavy particles (of like charge) interact via a two-body
repulsive Yukawa potential with all three and higher particle
interactions vanishing. It is also instructive to compute the charge
density of light quarks, \beq \label{density} \langle \bar{\psi}
\gamma_2\psi(x)\rangle_{\{x_i, q_i\}}=
-\frac{1}{\pi}\sum_i q_i V(x-x_i) 
\eeq It is clear from (\ref{density}) that each heavy
quark of charge $-q$ is surrounded by a cloud of light quarks with a
radius of roughly $\omega^{-1}$. The cloud has total charge $q$ that
screens the Coulomb potential of the heavy quark producing a meson,
similar to the heavy-light mesons of $QCD$ (such as the $B$-meson).

We will be most interested in the expectation value of the chiral
condensate $\langle \bar{\psi}\frac{1+\gamma^5}{2}\psi\rangle$. For
static sources this is given by\cite{Zahed} (we sketch the
calculation in the appendix), \beq
\label{ppcharge}\langle\bar{\psi}\frac{1+\gamma^5}{2}\psi(x)\rangle_{\{x_i,q_i\}}
=\frac{1}{2}e^{i\theta}\langle \bar{\psi}\psi\rangle_{L_1,L_2}
\prod_i \left(U(x-x_i)\right)^{-q_i}
\eeq where \beq \label{U}U(x) = \exp(2 \pi i  V'(x)) \stackrel{L_1
\to \infty}{=} \exp(-\pi i sgn(x) e^{-\omega |x|}) \eeq
\begin{figure}[t]
\begin{center}
\includegraphics[angle=-90, width = 0.45\textwidth]{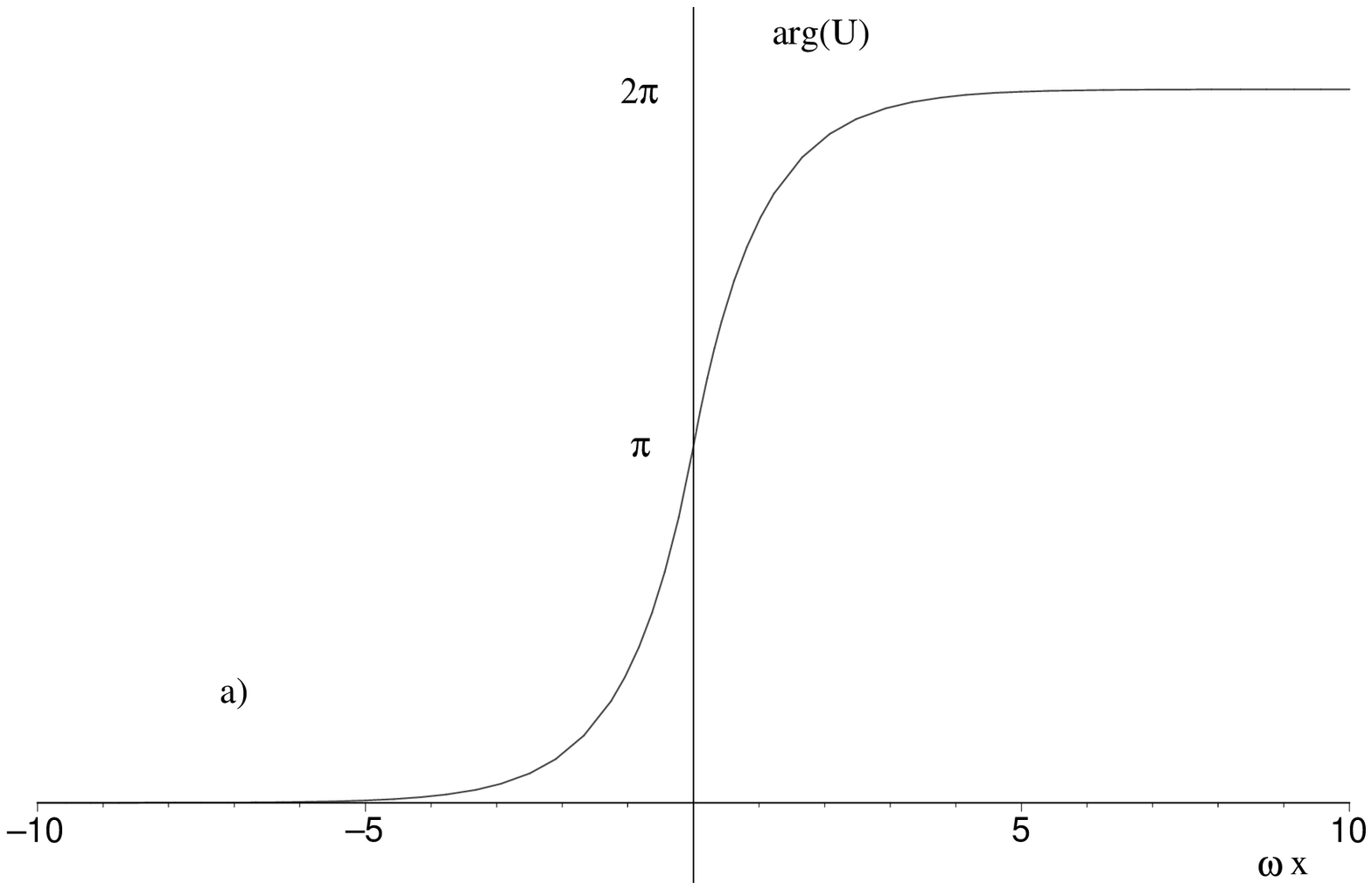}
\includegraphics[angle=-90, width = 0.45\textwidth]{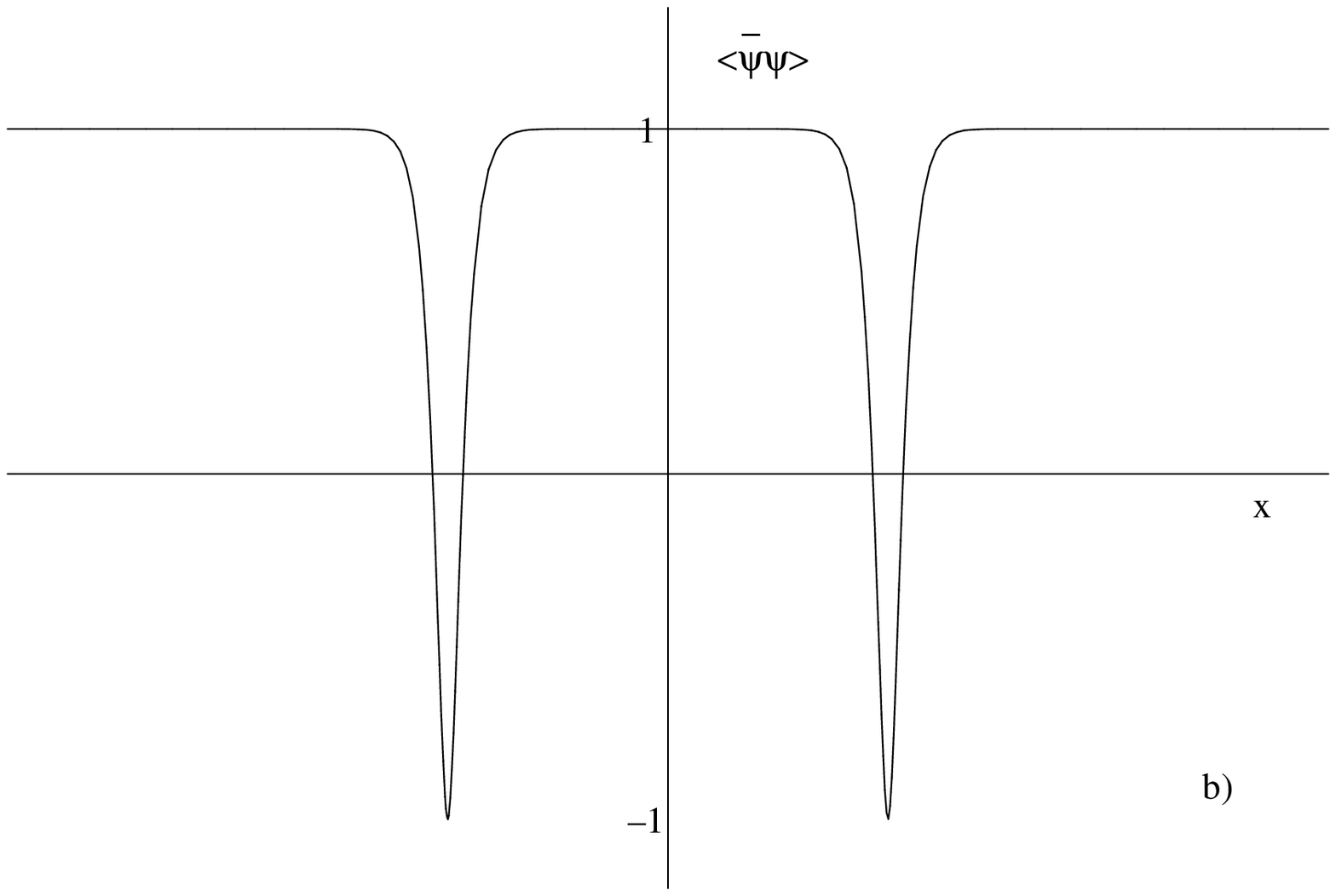}
\caption{a) The phase of the chiral condensate $\arg U(x) = \arg
\langle\bar{\psi}\frac{1+\gamma^5}{2} \psi(x)\rangle$ in the
background of a static unit charge placed at $x=0$ for $\theta = 0$.
b) The chiral condensate in units of
$\langle\bar{\psi}\psi\rangle_T$ in the background of two widely
separated static unit charges for $\theta = 0$.}  \label{FU}
\end{center}
\end{figure}
Thus, the introduction of static charges only affects the phase of
the chiral condensate. Moreover, $U(x) \to 1$ for $|x| \gg
\omega^{-1}$, so each static charge affects the chiral condensate
only in a region of radius roughly $\omega^{-1}$ - the size of the
meson. Notice that $U(x)$ makes one loop on the unit circle in the
complex plane as $x$ winds around the spatial circle (see Fig.
\ref{FU} a)). Thus, the phase of the condensate (\ref{ppcharge})
winds by $2 \pi N$ as $x$ moves around the spatial circle, where $N
= -\sum_i q_i$ is the total charge of the light fermions. So, the
total winding number of
$\langle\bar{\psi}\frac{1+\gamma^5}{2}\psi(x)\rangle_{\{x_i,q_i\}}$
is independent of the positions of the heavy quarks and, in fact, is
the same as for the model with the uniform background charge density
(\ref{osc}). However, the winding occurs in the vicinity of the
heavy charges, over the radius of each meson, as opposed to the
uniform background case, where the winding is uniformly smeared
across the whole system.\footnote{Technically, such a local nature
of the result comes from a non-trivial cancelation between
oscillating factors $e^{2 \pi i \rho x}$ originating from
integration over different modes in the path integral (see
appendix).} We expect this difference to be particularly important
in the dilute limit $\rho \ll \omega$ when the distance between
mesons is much larger than their size. In this regime, each meson
keeps its individual features and
$\langle\bar{\psi}\psi\rangle_{\{x_i, -q\}} \to
\langle\bar{\psi}\psi\rangle_{T}\cos(\theta)$ in the wide regions
between the mesons (see Fig. \ref{FU} b)). Here,
$\langle\bar{\psi}\psi\rangle_{T}$ is the infinite volume limit of
the chiral condensate at zero density and $\theta$ parameter and
finite temperature $T$. Thus, the uniform background approximation
is expected to fail badly in the dilute regime.

It is instructive to see what happens to the chiral condensate if we
arrange our heavy charges into a regular lattice, $x_j = j a$, $q_j
= - q$. Using (\ref{ppcharge}) and taking $L_1 \to \infty$, \beq
\langle\bar{\psi}\frac{1+\gamma^5}{2}\psi(x)\rangle_{\{x_i,q_i\}} =
\frac{1}{2} e^{i \theta} \langle \bar{\psi}\psi\rangle_T\, e^{i
\phi(x)}\eeq where $\phi(x)$ is a periodic function with period $x =
a$ and, \beq \phi(x) = \pi q
\frac{\sinh(\omega(x-a/2))}{\sinh(\omega a/2)}, \quad 0<x<a\eeq For
$\omega a \gg 1$ we have a crystal of widely spaced mesons much like
on Fig. \ref{FU} b).  In the high density limit, $\omega a \ll 1$,
the screening clouds of heavy charges overlap and the individual
mesons are washed out. Instead, we may approximate, \beq \phi(x)
\approx 2 \pi q \, \left(\frac{x}{a}-\frac{1}{2}\right), \quad 0< x
< a\eeq so that, \beq\label{osc2}
\langle\bar{\psi}\frac{1+\gamma^5}{2}\psi(x)\rangle_{\{x_i,q_i\}} =
\frac{1}{2} e^{i \theta} (-1)^{q} e^{2 \pi i \rho x} \langle
\bar{\psi}\psi\rangle_T\eeq where $\rho = q/a=N/L$. Thus, in this
limit we recover the uniform background approximation (\ref{osc}).
However, note that the existence of coherent, long-range,
oscillations of the chiral condensate (\ref{osc2}) is possible only
if the dynamics governing the heavy charges are such that they
organize a crystal-like state. For a one-dimensional statistical
system interacting with a finite range potential (\ref{V}) a true
crystal cannot form. However, the system may exhibit crystal order
on some finite distance scale $d \gg \omega^{-1}$. In this case, we
expect that the plane wave behaviour of the chiral condensate will
also persist on the same distance scale $d$. Otherwise, if the
mesons form a disordered, weakly-interacting gas the oscillations
(\ref{osc2}) will be washed out on distance scales $x \gg
\omega^{-1}$, as we shall show shortly.

In fact, equation (\ref{ppcharge}) suggests that the external
charges act as impurities, whose effect is to disorder the chiral
condensate. If the impurities are in a weakly interacting regime,
the disorder is ``random". This is precisely the situation that we
will analyze in the next section.

\subsection{Statistical Model}
Let us now make the heavy charges dynamical and analyze the
statistical model (\ref{eff}). 
Our main objective is to compute the chiral condensate
(\ref{O}),(\ref{ppcharge}), \bea \langle
\bar{\psi}\frac{1+\gamma^5}{2}\psi(x)\rangle &=& \frac{1}{2}
e^{i\theta} \langle \bar{\psi}\psi\rangle_T
\frac{1}{Z}\sum_{n=0}^{\infty} \frac{z^n}{n!} \int dx_1..dx_n
\prod_i U(x-x_i)^{q} \exp(-\frac{\beta}{2} \sum_{ij} q^2 e^2 V(x_i -
x_j))\nn\\\label{Ug}&=& \frac{1}{2} e^{i\theta} \langle
\bar{\psi}\psi\rangle_T \sum_{n=0}^{\infty} \frac{1}{n!}\int
dx_1..dx_n \prod_i (U(x-x_i)^{q}-1) g_n(x_1,..x_n)\eea where
$g_n(x_1,..x_n)$ is the $n$-point correlation function, \beq
g_n(x_1,..x_n) = z^n \frac{1}{Z}\sum_{m=0}^{\infty} \frac{z^m}{m!}
\int dx_{n+1}..dx_{n+m} \exp(-\frac{\beta}{2} \sum_{ij=1}^{n+m} q^2
e^2 V(x_i - x_j))\eeq Notice that the chiral condensate is sensitive
only to short-distance properties of the correlation functions
$g_n(x_1..x_n)$ as the range of $U(x)^q - 1$ is roughly
$\omega^{-1}$.

We would like to perform the Meyer expansion in activity $z$. The
leading term in the equation of state, as always, is, \beq
\label{ideal}\beta P = z^*\eeq where $P$ is the pressure and $z^* =
z \exp(-\frac12 {\beta}q^2 e^2 V(0))$ ($z^*$ includes the
self-interaction energy of each meson). Then, \beq \rho_h = z
\frac{\d}{\d z} (\beta P) = z^*\eeq where $\rho_h$ is the density of
heavy particles, \beq q \rho_h = \rho\eeq As the range of the
potential $V$ is $\omega^{-1}$ and strength $e^2 V \sim \omega$, the
corrections to the equation of state for $T \sim \omega$ are
suppressed in the Meyer cluster expansion by powers of $z^*/\omega$.
So for, $\rho \ll \omega$, $T \sim \omega$ our system behaves like a
weakly-interacting gas of mesons. Moreover, in this regime, at
leading order the $n$ point function on distances $x \sim
\omega^{-1}$ scales as ${z^*}^{n}$ so that terms in (\ref{Ug})
involving  $g_n(x_1,..x_n)$ are suppressed by $(z^*/\omega)^n$. The
leading correction to the chiral condensate comes from the $n=1$
term. Recalling, $g_1(x) = \rho_h$, \beq \langle
\bar{\psi}\frac{1+\gamma^5}{2}\psi\rangle \approx \frac{1}{2}
e^{i\theta} \langle \bar{\psi}\psi\rangle_T (1 + \rho_h \int dx
(U(x)^q -1)) \eeq Performing the integral, \beq \label{ppn}\langle
\bar{\psi}\psi\rangle \approx  (1 - \frac{\rho_h}{\omega} u(q))
\langle{\bar{\psi}\psi}\rangle_T \cos(\theta) \eeq where, \beq u(q)
= 2 \int_0^{\pi q} dt \frac{(1-\cos(t))}{t} = 2 (\log(\pi q) +
\gamma - Ci(\pi q))\eeq Thus, we have calculated the first
correction in density to the chiral condensate in the regime of a
weakly interacting dilute meson gas. The result (\ref{ppn}) is
reminiscent of the behaviour of the chiral condensate in $N_c = 2$
$QCD$ at small baryon density\cite{KSTVZ}, in $N_c = 3$ $QCD$ at
small isospin density\cite{SS} and in the ``dilute" nuclear
matter\cite{nuclear}. In all of these theories, one thinks of the
system as being composed of a dilute gas of particles $M$ (diquarks
for $N_c = 2$ $QCD$, pions for $N_c = 3$ $QCD$, nuclons for nuclear
matter) and obtains, \beq \langle \bar{\psi} \psi\rangle_\rho
\approx \langle \bar{\psi} \psi\rangle_0 + \rho \langle
M|\bar{\psi}\psi|M\rangle \eeq where $|M\rangle$ is a one particle
state in vacuum (with normalization $\langle M(p)|M(p')\rangle = (2
\pi)^d \delta^d(p-p')$). So, we identify, \beq \label{ppq}\langle
q|\bar{\psi} \psi| q \rangle_{T,\theta} = \int dx (U(x)^q-1)
\,\langle \bar{\psi} \psi\rangle_T \cos(\theta) = - \frac{1}{\omega}
u(q) \langle \bar{\psi} \psi\rangle_T \cos(\theta)\eeq where
$|q\rangle$ denotes our heavy-light meson state.

So far we have concentrated  on the region $T \sim \omega$ where the
criterion for the applicability of Meyer's expansion was $\rho \ll
\omega$. Now, let's analyze the low temperature regime $T \ll
\omega$. In this case, the interaction effectively becomes hardcore
of range $\omega^{-1} \log(\omega/T)$, so for the Meyer expansion to
be valid, we need $z^* \ll \omega/\log(\omega/T)$. If this condition
is satisfied, the chiral condensate at leading order in $\rho$ is
again given by (\ref{ppn}). Moreover, we actually expect the
expressions (\ref{ppn}), (\ref{ppq}) to remain valid in a wider
range $\rho \ll \omega$ down to the extreme quantum regime at $T =
0$, based on general phase-space arguments.

Finally, let's study the high-temperature regime $T \gg \omega$. In
this case, it can be shown that the corrections to the ideal gas
equation of state (\ref{ideal}) are suppressed by powers of $z/T$.
In particular, if $\omega \ll \rho \ll T$ we are still in the weakly
interacting (but not dilute!) regime. In this case it is convenient
to rewrite (\ref{Ug}) as, \beq \langle
\bar{\psi}\frac{1+\gamma^5}{2}\psi(x)\rangle = \frac{1}{2}
e^{i\theta} \langle \bar{\psi}\psi\rangle_T
\exp\left(\sum_{n=1}^{\infty} \frac{1}{n!}\int dx_1..dx_n \prod_i
(U(x-x_i)^{q}-1) g_n(x_1,..x_n)_{conn}\right)\eeq where
$g_n(x_1,..x_n)_{conn}$ denotes the fully connected $n$-point
correlation function. It is easy to show that the terms in the
exponent involving the $n$-point correlation function are suppressed
by $(z/T)^{n-1}$ compared to the leading term and, \beq
\label{ppT}\langle \bar{\psi} \psi\rangle \approx
\exp\left(-\frac{\rho_h}{\omega} u(q)\right)\langle \bar{\psi}
\psi\rangle_T \cos(\theta) \eeq The expression (\ref{ppT}) agrees
with (\ref{ppn}) in the dilute limit $\rho \ll \omega$. In the dense
gas limit, $\omega \ll \rho \ll T$, the chiral condensate
exponentially decreases with density. Note that for $T \gg \omega$
the chiral condensate at zero density is already exponentially
suppressed with temperature compared to $T = 0$ (see eq.
(\ref{ppexplicit})).

\subsection{Correlation Functions}
To answer the question of whether any remnants of the oscillating
behaviour (\ref{osc}) exist in our model, the computation of the
chiral condensate presented in the previous section is not
sufficient. Indeed, translational invariance implies that the chiral
condensate is uniform. Instead, we must compute the static
correlation function, $\langle S_+(x) S_-(y)\rangle$, $S_{\pm}(x)
=\bar{\psi} \frac{1\pm\gamma_5}{2}\psi(x)$. We would like to see on
what distance scales this correlation function exhibits plane wave
structure (\ref{osc}). Integrating out light degrees of freedom,
\beq \label{SScorr}\langle S_+(x)S_-(y)\rangle_{\{x_i, q_i\}} =
S(x-y) \prod_i (U(x-x_i))^{-q_i} (U(y-x_i))^{q_i}\eeq where, \bea
S(x-y) &=& \langle S_+(x)S_-(y)\rangle_{L_1,L_2} =
\frac{1}{4}|\langle\bar{\psi}\psi\rangle_{L_1,L_2}|^2 e^{4 \pi
G_\omega(x-y)}\\\quad G_\omega(x) &=&  \frac{1}{L_1 L_2} \sum_{p}
\frac{1}{p^2+\omega^2}e^{i p x},\quad p = \left(\frac{2 \pi
m_1}{L_1}, \frac{2 \pi m_2}{L_2}\right),\,m_1,m_2 \in {\mathbb
Z}\eea Thus, \beq \label{SpSm}\langle S_+(x)S_-(y)\rangle = S(x-y)
F(x-y) \eeq \beq \label{F} F(x-y) = \exp\left(\sum_{n=1}^{\infty}
\frac{1}{n!}\int dx_1..dx_n \prod_i (U(x-x_i)^q U(y-x_i)^{-q}-1)
g_n(x_1,..x_n)_{conn}\right)\eeq So the correlation function
factorizes into two pieces. The first, $S(x-y)$, is just the
correlation function in vacuum. The second, $F(x-y)$, contains the
finite density information.

\begin{figure}[t]
\begin{center}
\includegraphics[angle=-90, width = 0.45\textwidth]{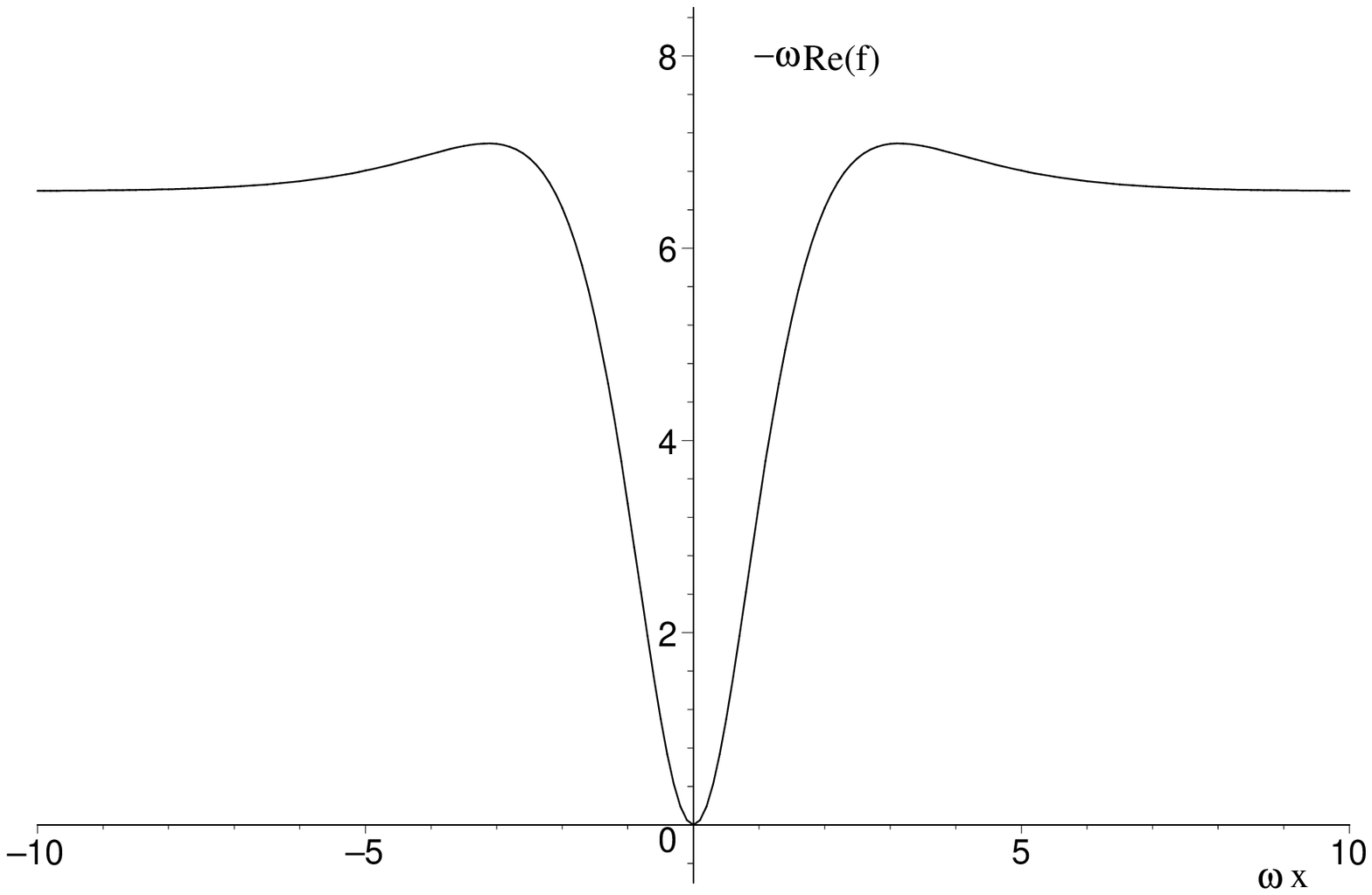}
\includegraphics[angle=-90, width = 0.45\textwidth]{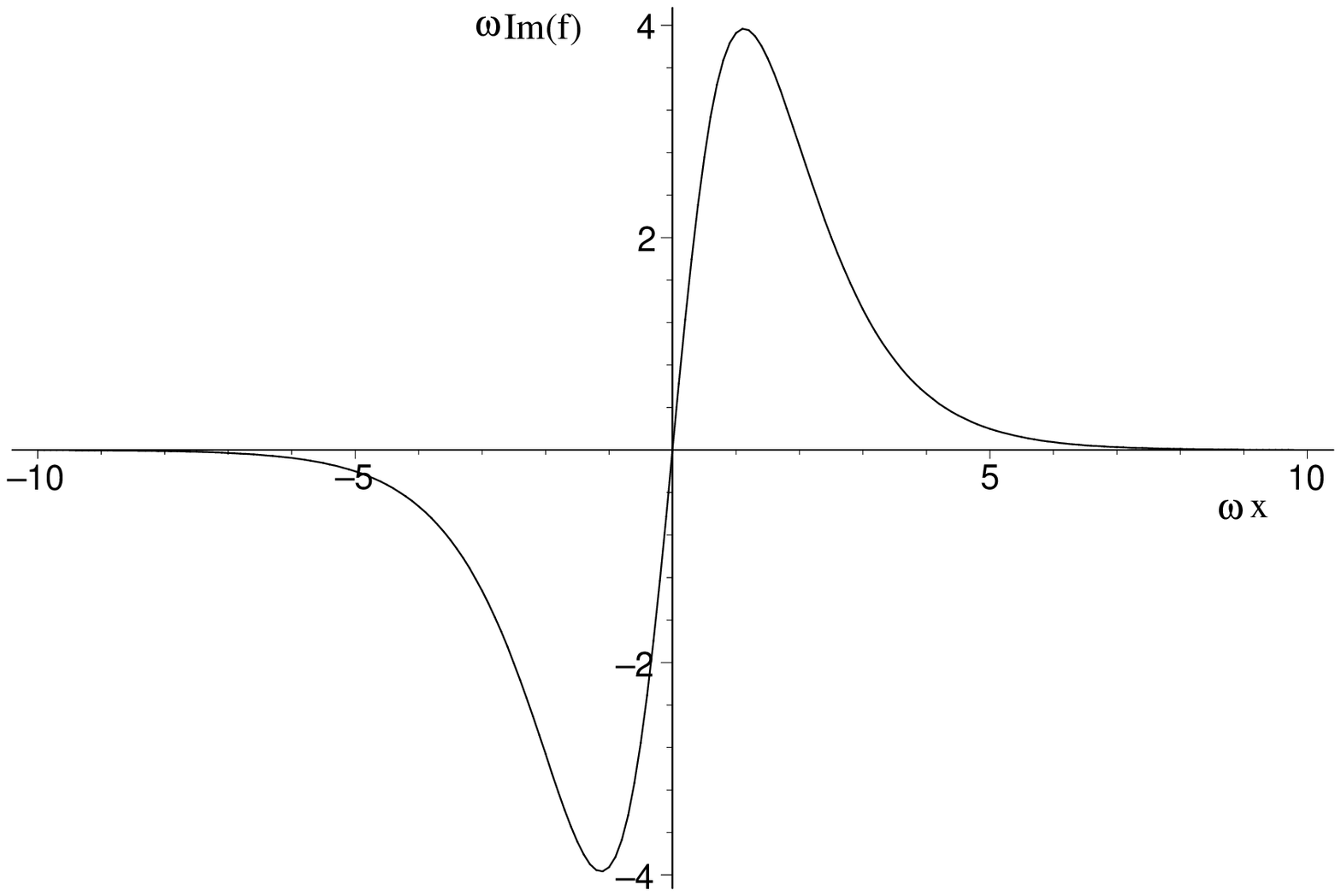}
\caption{Function $f(x)$ that enters the correlator $\langle S_+(x)
S_-(0)\rangle$ (see eqs. (\ref{SpSm}), (\ref{F})). Here we chose
$q=1$.} \label{profile}
\end{center}
\end{figure}

As noted in the previous section, in the regime where the Meyer
expansion is applicable, we may truncate the series in the exponent
of (\ref{F}) at the leading ($n$ = 1) term, \beq \label{F} F(x-y) =
e^{{\rho_h} f(x-y)}\eeq where, \beq f(x-y) = \int dx_1 (U(x-x_1)^q
U(y-x_1)^{-q}-1)\eeq The function $f(x)$ is plotted in Fig.
\ref{profile}.
It is easy to see that for $|x-y| \gg \omega^{-1}$, \beq f(x-y) \to
\int dx_1 (U(x-x_1)^q - 1) + \int dx_1 (U(y-x_1)^{-q} -1) =
-\frac{2}{\omega} u(q)\eeq Thus,  $\langle S_+(x) S_-(y)\rangle$
will not exhibit oscillations (\ref{osc}) for $|x-y| \gg
\omega^{-1}$.

Note that $S(x-y) \to \frac{1}{4}|\langle\bar{\psi}\psi\rangle_T|^2$
for $|x-y| \to \infty$ so, \beq \langle S_+(x)
S_-(y)\rangle\stackrel{|x-y|\to \infty}{=}
\frac{1}{4}|\langle\bar{\psi}\psi\rangle_T|^2 \exp(-2
\frac{\rho_h}{\omega} u(q)) = \langle S_+\rangle \langle
S_-\rangle\eeq and the correlation function clusters.

We may also investigate the short distance behaviour. Expanding
$f(x)$ in a Taylor series in $\omega x$, \beq  f(x) =
\frac{1}{\omega} (2 \pi i q \omega x - \frac{1}{2} \pi^2 (q \omega
x)^2 + ...)\eeq Hence, for $|x| \ll \omega^{-1}$, \beq \label{osc3}
\langle S_+(x) S_-(0)\rangle \approx \exp(2 \pi i \rho x -
\frac{\pi^2 q}{2} \frac{\rho}{\omega} (\omega x)^2) S(x)\eeq The
above equation clearly exhibits the plane wave behaviour with period
$x = \rho^{-1}$. However, recall that eq. (\ref{osc3}). is valid
only for $|x| \ll \omega^{-1}$. Thus, in the dilute limit $\rho \ll
\omega$, no full oscillations appear and, in fact, eqs.
(\ref{SpSm}),(\ref{F}) are more appropriately written as, \beq
\langle S_+(x) S_-(0)\rangle = (1 + \rho_h f(x)) S(x)\eeq In the
dense gas regime, $\rho \gg \omega$, the oscillations are, indeed,
present on short distance scales, however, as eq. (\ref{osc3})
shows, they become damped for $x \gtrsim
(\frac{\omega}{\rho})^\frac12 \omega^{-1}$ and disappear altogether
for $x \gg \omega^{-1}$. Moreover, these oscillations modulate the
zero-density correlator $S(x)$, which itself has a quite non-trivial
behaviour for distances $x \lesssim \omega^{-1}$.

\section{Conclusion}

In this paper we have analyzed some puzzles related to the Schwinger
model at finite density. We have shown that the well-known
plane-wave behaviour of the chiral condensate is a consequence of
explicit breaking of translational invariance by a background charge
density. Similarly to the non-conservation of axial charge, the
non-conservation of total momentum is globally saturated in sectors
of non-trivial topological charge. In fact, the breaking of
translational symmetry at finite density is a much simpler
phenomenon than the breaking of chiral symmetry by the anomaly as
the former appears already on the classical level, while the later
is a purely quantum phenomenon.

In the second part of this paper, we have explored the question:
``What happens if the uniform background density is replaced by a
dynamical, but heavy, field?" To answer this question, we have
analyzed a statistical model in which the heavy neutralizing charge
comes from an ensemble of classical particles. We have shown that
the effect of heavy charges is to disorder the chiral condensate. In
the regime where the gas of heavy charges is almost ideal, the
chiral condensate is spatially uniform and decreasing with density.
For the charge density $\rho \ll \omega$ the ``disorder" is weak and
we compute the leading density correction to the chiral condensate
(\ref{ppn}). In the dense gas regime $\rho \gg \omega$ the disorder
leads to an exponential suppression of the chiral condensate. In
both of these regimes, the condensate does not exhibit any
oscillations on distance scales $x \gg \omega^{-1}$, as is clear
from computing the correlator $\langle S_+(x) S_-(0)\rangle  =
\langle \bar{\psi}\frac{1+\gamma^5}{2} \psi(x)
\bar{\psi}\frac{1-\gamma^5}{2} \psi(0)\rangle$. The only remnant of
oscillatory behaviour comes at high density $\rho \gg \omega$ in the
short distance behavior of $\langle S_+(x) S_-(0)\rangle$ for $x \ll
\omega^{-1}$.

In fact, we have argued that the only way for the oscillations to
survive on distance scales larger than $\omega^{-1}$ is for the
system to be in the high density regime $\rho \gg \omega$ and the
heavy charges to crystalize. In the dilute regime, $\rho \ll \omega$
we do not expect to recover the oscillatory behaviour even if the
heavy charges were to crystalize.

So clearly the uniform background approximation generically does not
accurately model the situation where the neutralizing charge is
dynamical, rendering the results of \cite{Susskind,Kao,Schaposnik}
unphysical. Indeed, we expect such an approximation to work well if
the light and heavy neutralizing charges are largely decoupled from
each other (e.g. valence electrons in a metal). However, in the
Schwinger model the light fermions are very strongly coupled to the
neutralizing charges producing
heavy-light mesons. 
The approximation fails particularly badly in the dilute phase,
where the distance between the mesons is much larger than their
size. To apply the uniform background charge approximation here
would be akin to treating the nucleii in a dilute atomic gas as
uniform.

We conclude by noting that we certainly have not analyzed the entire
phase-diagram of the two-flavour heavy-light Schwinger model. We
have treated the gas of heavy-light mesons classically and have not
touched upon the quantum regime at all. Neither have we analyzed the
regime where the classical system is far from an ideal gas limit and
the interactions between mesons are important. These regimes are
subject to further investigation and, in fact, have a higher chance
of exhibiting the plane-wave behaviour of chiral condensate on
distance scales larger than $\omega^{-1}$ than the ``random
disorder" case considered here.

\section*{Acknowledgements}
I would like to thank A. Zhitnitsky, M. Stephanov, M. Forbes, I.
Klebanov and S. Sachdev for useful discussions. This work
 was supported, in part, by the Natural Sciences and Engineering
Research Council of Canada.

\appendix
\section{Explicit Calculation of the Chiral Condensate at Finite
Density} The purpose of this appendix is to perform the calculation
of the partition function and chiral condensate at finite background
charge density on a Euclidean torus. In case when the background
charge is in the form of discrete charges (Wilson loops) this
computation has been performed before\cite{Gordon, Zahed}. For a
uniform background charge density, the calculation has been done on
an infinite Euclidean plane\cite{Schaposnik}. Here, we keep the size
of our torus finite throughout the calculation, gaining complete
control of all the infra-red subtleties. For a detailed study of the
Schwinger model on the torus (at zero density) see \cite{Azakov}.

We work in a gauge where the (fermion) gauge fields are (anti)
periodic in the temporal direction, with the transition functions
(\ref{anti}), \beq V_1(x_1) = -1, \quad V_2(x_2) = e^{2\pi i n
x_2/L_2}\eeq We decompose the gauge fields as, \beq A_{\mu} =
t_{\mu} + \d_{\mu} \alpha + \epsilon_{\mu \nu} \d_{\nu} b +
A^{n}_{\mu}\eeq where $\alpha$ and $b$ are both periodic fields on
the torus orthogonal to unity ($\int d^2 x \,\alpha = \int d^2 x \,b
= 0$). The variable $t_{\mu}$ is the so-called toron field and plays
a crucial part in all the calculations. $t_{\mu}$ is effectively an
angular variable, with $t_{\mu} \sim  t_{\mu} + \frac{2 \pi m}{e
L_{\mu}}, \, m \in {\mathbb Z}$. We shall consider only the case
where the total background charge is integral, $\int dx_1
j_2^{ext}(x_1) = - N, \, N \in {\mathbb Z}$ so that the angular
nature of $t_{\mu}$ is unspoiled. $A^n_{\mu}$ is the ``instanton"
field in the $n$-th topological sector. We choose, \beq A^n_1 =
0,\quad A^n_2 = \frac{2 \pi n x_1}{e L_1 L_2} \eeq which obeys the
periodicity conditions (\ref{bc1}),(\ref{bc2}).

First, let's compute the partition function, \beq Z = \int \, {\cal
D} A {\cal D}\bar{\psi} {\cal D} \psi \, e^{i e \int d^2
x\,j^{ext}_{2} A_2}\, e^{-S} e^{i n \theta}\eeq with the
normalization $Z = 1$ for $j^{ext}_2 = 0$. For vanishing mass, only
contributions from the trivial topological sector survive (recall
that there are precisely $|n|$ fermion zero modes in a sector with
topological charge $n$ with $\gamma_5 = sgn(n)$) and, \beq Z =
\int_{n=0} DA \,\det(\gamma_{\mu} D_{\mu}) e^{-\frac{1}{2} \int d^2
x F^2} e^{i e \int d^2 x\,j^{ext}_{2} A_2}\eeq The (regularized)
Dirac determinant is given by (see \cite{Azakov} and references
therein), \beq {\det}'(L_1 (\gamma_{\mu} D_{\mu})) = \det {\cal N}
\exp(\Gamma(b) + \Gamma(t,n))\eeq where $\det'$ denotes the
determinant with the zero mode contributions deleted and, \bea
\Gamma(b) &=& -\frac{1}{2}
\omega^2 \int d^2 x \, \d_{\mu} b \d_{\mu} b\\
\Gamma(t,n) &=& \delta_{n,0} \left(-2 \pi |\tau| \theta_1^2 +
\log(|\nu_2(\pi \theta \tau, \tau)|^2 \eta(\tau)^{-2})\right)  -
\frac{1}{2} |n| \log\left(\frac{2\, |n|}{|\tau|}\right)\eea Here,
$\theta = \theta_1 + i \theta_2 = \frac{e (t_1 + i t_2) L_1}{2\pi}$,
$\tau = i \frac{L_2}{L_1}$ and ${\cal N}$ is the matrix of zero mode
inner products, \bea {\cal
N}_{ij} &=& \int d^2 x \, \chi^\dagger_i(x) \chi_j (x)\\
\chi_i(x) &=& e^{i e \alpha(x)} e^{e \gamma_5 b(x)} \chi^0_i(x)\eea
where $\chi^0_i(x)$ are the orthonormal zero modes of the operator
$D_0 = \gamma_{\mu} (\d_{\mu} - i e (t_{\mu} + A^n_{\mu}))$. Thus,
\beq Z = \frac{\int dt_1 dt_2 \,e^{-i e N t_2 L_2}
e^{\Gamma(t,0)}}{\int dt_1 dt_2 \,e^{\Gamma(t,0)}}\, \frac{\int
{\cal D}b\, e^{i e \int d^2x\, \d_1 j_2^{ext}(x_1) \, b(x)}
e^{-S(b)}}{\int {\cal D}b\, e^{-S(b)}}\eeq where, \beq S(b) =
\frac{1}{2} \int d^2 x\,b(x) (-\d^2) (-\d^2 + \omega^2 ) b(x)\eeq
Performing the integral over the toron fields we obtain, \beq
\frac{\int dt_1 dt_2 \,e^{-i e N t_2 L_2} e^{\Gamma(t,0)}}{\int dt_1
dt_2 \,e^{\Gamma(t,0)}} = \exp(-L_2 \frac{\pi N^2}{2 L_1})
\label{dt}\eeq we recognize $\epsilon_F = \frac{\pi N^2}{2 L_1}$ as
the Fermi energy of free massless Dirac fermions in $2d$ at fermion
number $N$.

The integration over $b$ field gives, \beq \frac{\int {\cal D} b
e^{\int \eta(x) b(x)} e^{-S(b)}}{\int {\cal D} b  e^{-S(b)}} =
\exp\left(\frac{1}{2}\int d^2 x d^2 y \,\eta(x) G(x-y)
\eta(y)\right)\label{gaussb}\eeq with the propagator, \bea G(x) &=&
\frac{1}{\omega^2} \left(\bar{G}_0(x) -
\bar{G}_\omega(x)\right)\\\bar{G}_{\lambda}(x) &=& \frac{1}{L_1
L_2}\sum_{p \neq 0}\frac{1}{p^2+\lambda^2}e^{i p x},\quad p =
\left(\frac{2 \pi m_1}{L_1}, \frac{2 \pi m_2}{L_2}\right), \,
m_1,m_2 \in {\mathbb Z} \eea

So we find, \beq \frac{\int {\cal D}b\, e^{i e \int d^2x\, \d_1
j_2^{ext}(x_1) \, b(x)} e^{-S(b)}}{\int {\cal D}b\, e^{-S(b)}} =
\exp\left(-L_2 \frac{e^2}{2} \int dx dy \,j^{ext}_2(x)(V(x-y) -
\frac{1}{L_1 \omega^2})j^{ext}_2(y)\right)\label{Db}\eeq with $V(x)$
given by (\ref{V}). Combining the global and local pieces
(\ref{dt}), (\ref{Db}), \beq Z = \exp\left(-L_2 \frac{e^2}{2} \int
dx dy \,j^{ext}_2(x) V(x-y) j^{ext}_2(y)\right)\label{appZ}\eeq

For a uniform background charge density, $j^2_{ext} = -
\frac{N}{L_1}$ the contribution to (\ref{appZ}) comes only from the
global piece and is given by (\ref{dt}). On the other hand for
discrete integral charges, $j^2_{ext}(x) = \sum_i q_i \delta(x-x_i)$
and, \beq Z = \langle \prod_i W(x_i,q_i) \rangle = \exp\left(-L_2
\frac{1}{2} \sum_{i j} q_i q_j e^2 V(x_i - x_j)\right)\eeq

Now, let's compute the chiral condensate, $\langle
\bar{\psi}\frac{1+\gamma_5}{2}\psi(x)\rangle$. For $m=0$, it
receives a contribution only from the topological sector with $n=1$,
\beq \langle \bar{\psi}\frac{1+\gamma_5}{2}\psi(x)\rangle =  Z^{-1}
\int_{n=1} DA\, L_1 \xi^{\dagger}(x) \frac{1+\gamma_5}{2}\xi(x) \,
{\det}'(L_1 \gamma_{\mu} D_{\mu})e^{i \theta} e^{-\frac{1}{2} \int
d^2 x' F^2} e^{i e \int d^2 x'\, j^{ext}_2(x'_1) A_2(x')} \eeq where
$\xi$ is the normalized zero mode of the operator $\gamma_{\mu}
D_{\mu}$. We have, \beq \xi(x) = {\cal N}^{-\frac12} \chi(x) = {\cal
N}^{-\frac12} e^{i e \alpha(x)} e^{e \gamma_5 b(x)}\chi^0(x) \eeq As
noted, $\chi^0$ is the normalized zero mode of the Dirac operator in
the background of toron and instanton fields, \beq
\gamma_{\mu}(\d_{\mu}-i e (t_{\mu} + A^n_{\mu}))\chi^0(x) = 0 \eeq
obeying the boundary conditions (\ref{bc1}),(\ref{bc2}). In the
$n=1$ sector we have a single zero mode, \beq \chi^0(x_1,x_2) =
\frac{1}{L_1}\left(\frac{2}{|\tau|}\right)^{\frac14} e^{-\pi |\tau|
\theta^2_1} e^{\frac{\pi}{|\tau|}(2i
\tilde{x}_1\tilde{x}_2-{\tilde{x}^2}_2) + 2 \pi \tilde{x}_2 \theta}
\nu_4(\pi (\tilde{x}_1 + i\tilde{x}_2)-\pi \theta \tau|\tau)
\xi_+\eeq where $\tilde{x}_i = x_i/L_1$ and $\xi_+$ is the spinor
with positive chirality $\gamma_5 \xi_+ = \xi_+$.

Thus, \bea \langle \bar{\psi}\frac{1+\gamma_5}{2}\psi(x)\rangle &=&
e^{i \theta} e^{-S_0} e^{\frac{2 \pi i}{L_1} \int dx_1 x_1
j^{ext}_2(x_1)}\,\frac{\int dt_1 dt_2 \,L_1 \,
\chi^{\dagger}_0\chi_0(x) e^{-i e N t_2 L_2} e^{\Gamma(t,1)} }{\int
dt_1 dt_2 e^{-i e N t_2 L_2} e^{\Gamma(t,0)}}\times\nn\\&&
\frac{\int {\cal D} b \,e^{2 e b(x)} e^{ie \int d^2 x' \d_1
j_2(x'_1)\,b(x')} e^{-S(b)}}{\int {\cal D} b \,e^{ie \int d^2 x'
\d_1 j_2(x'_1)\,b(x')} e^{-S(b)}}\eea where $S_0 = \frac{2
\pi^2}{e^2 L_1 L_2}$ is the ``bare" instanton action. Performing the
average over the toron fields, \beq \frac{\int dt_1 dt_2 \,L_1 \,
\chi^{\dagger}_0\chi_0(x) e^{-i e N t_2 L_2} e^{\Gamma(t,1)} }{\int
dt_1 dt_2 e^{-i e N t_2 L_2} e^{\Gamma(t,0)}} = \frac{1}{L_1}
\eta^2(\tau) (-1)^N e^{2 \pi i N x_1/L_1}\label{dtpsi}\eeq Taking
the average over the $b$ field with the help of (\ref{gaussb}), \bea
&& \frac{\int {\cal D} b \,e^{2 e b(x)} e^{ie \int d^2 x' \d_1
j_2(x'_1)\,b(x')} e^{-S(b)}}{\int {\cal D} b \,e^{ie \int d^2 x'
\d_1 j_2(x'_1)\,b(x')} e^{-S(b)}} = e^{2 e^2 G(0)} \exp\left(2 i e^2
\int d^2 x'\, G(x-x') \d_1 j^{ext}_2(x_1')\right)=\nn\\ && =e^{2 e^2
G(0)} \exp\left(2 \pi i \int dx'_1
\left(\frac{x_1-x'_1}{L_1}-\frac{1}{2}sgn(x_1-x'_1)-V'(x_1-x'_1)\right)
j^{ext}_2(x'_1)\right)\label{dbpsi}\eea Now, combining eqs.
(\ref{dtpsi}),(\ref{dbpsi}), \beq \langle
\bar{\psi}\frac{1+\gamma_5}{2}\psi(x)\rangle=\frac{1}{2}e^{i\theta}(-1)^N
\langle\bar{\psi}\psi\rangle_{L_1,L_2} \exp\left(-2\pi i\int
dx'_1\left(\frac{1}{2}
sgn(x_1-x'_1)+V'(x_1-x'_1)\right)j^{ext}_2(x_1')\right)\eeq where
the chiral condensate in the absence of external charge and at
$\theta = 0$ is given by, \beq \label{ppexplicit}\langle
\bar{\psi}\psi\rangle_{L_1,L_2} = \frac{2}{L_1} e^{-S_0}
\eta^2(\tau) e^{2 e^2 G(0)}=\lim_{x\to 0} \frac{1}{\pi|x|}e^{-2\pi
G_{\omega}(x)} \to \left\{\begin{array}{ll} \frac{\omega}{2\pi}
e^{\gamma} & \quad T \to 0,\,
L_1 \to \infty\\
2 T e^{-\frac{\pi T}{\omega}} & \quad T \to \infty,\, L_1 \to
\infty\end{array}\right. \eeq If the background charge density is
uniform, \beq \langle \bar{\psi}\frac{1+\gamma_5}{2}\psi(x)\rangle
=\frac{1}{2} e^{i \theta} e^{2 \pi i N x_1/L_1}\langle
\bar{\psi}\psi\rangle_{L_1,L_2}\eeq In this case the oscillating
factor comes solely from the integration over the toron fields
(\ref{dtpsi}). On the other hand, if the external charges are
discrete and integral, \beq \langle
\bar{\psi}\frac{1+\gamma_5}{2}\psi(x)\rangle = \frac{1}{Z}\langle
\bar{\psi} \frac{1+\gamma_5}{2}\psi(x)\prod_i W(x_i,q_i)\rangle =
\frac{1}{2} e^{i\theta} \exp\left(-2 \pi i \sum_i q_i
V'(x-x_i)\right) \langle\bar{\psi}\psi\rangle_{L_1,L_2}\eeq and the
long range oscillating factor is canceled between the global
(\ref{dtpsi}) and local (\ref{dbpsi}) parts.

A similar computation can be performed to obtain the result
(\ref{SScorr}) for the correlation function of chiral densities.




\end{document}